\documentclass[twocolumn,showpacs,aps,prd,superscriptaddress,floatfix]{revtex4}
\usepackage{graphicx}
\usepackage{dcolumn}
\usepackage{amsmath}
\usepackage{epsfig}

\newcommand{\BABARPubYear}    {06}
\newcommand{\BABARPubNumber}  {032}

\newcommand{\SLACPubNumber} {11902}
\newcommand{\LANLNumber} {0606031}

\input babarsym

\def\klong {\KL}
\def\kshort {\KS}
\def\kstar0 {\ensuremath{K^{*0}}\xspace}
\def\kstar {\ensuremath{K^{*}}\xspace}

\providecommand{\BBbar}{\BB}

\providecommand{\DE}{\DeltaE}
\providecommand{\calB}{{\ensuremath{\cal B}}\xspace}

\providecommand{\bfemsix}{\ensuremath{\calB(10^{-6})}}

\def\m2ksks{\ensuremath{m^2_{\KS\KS}}\xspace}

\providecommand{\kskskl}{\ensuremath{\kshort\kshort\KL}\xspace}

\providecommand{\ksksks}{\ensuremath{\kshort\kshort\kshort}\xspace}

\providecommand{\phiks}{\ensuremath{\phi\kshort}\xspace}

\providecommand{\KKPiZ}{\ensuremath{\kshort\kshort\piz}\xspace}
\providecommand{\KKKst}{\ensuremath{KK\kstar}\xspace}

\providecommand{\bigTableStretch}{\renewcommand\arraystretch{1.3}}

\newcommand\dbline{\noalign{\vskip 0.10truecm\hrule}\noalign{\vskip 2pt}%
  \noalign{\hrule\vskip 0.10truecm}}
\newcommand\sgline{\noalign{\vskip 0.10truecm\hrule\vskip 0.10truecm}}

\providecommand{\MCeffKsKsKl}{\ensuremath{8.7}\xspace}

\providecommand{\effKL}{\ensuremath{96.1}\xspace}
\providecommand{\effKLSyst}{\ensuremath{8.0}\xspace}

\providecommand{\kskskslCorrEff}{\ensuremath{8.1}\xspace}
\providecommand{\ksksklEffProdBRs}{\ensuremath{3.8}\xspace}
\providecommand{\phiksklksCorrEff}{\ensuremath{5.6}\xspace}
\providecommand{\phiksklksEffProdBRs}{\ensuremath{0.91}\xspace}
\providecommand{\NsimEvtKsKsKl}{\ensuremath{5892}\xspace}

\providecommand{\NsimKKPiZ}{\ensuremath{2.3}\xspace}
\providecommand{\NsimKsKsKs}{\ensuremath{20}\xspace}
\providecommand{\NsimKKKst}{\ensuremath{70}\xspace}

\providecommand{\YBkgKsKsKl}{\ensuremath{5777\pm79}\xspace}
\providecommand{\YKsKsKl}{\ensuremath{23}\xspace}
\providecommand{\PErrKsKsKl}{\ensuremath{+23}\xspace}
\providecommand{\MErrKsKsKl}{\ensuremath{-22}\xspace}
\providecommand{\YBkgPhiKs}{\ensuremath{202\pm15}\xspace}
\providecommand{\Yphiksklks}{\ensuremath{8.3}\xspace}
\providecommand{\PErrphiksklks}{\ensuremath{+5.5}\xspace}
\providecommand{\MErrphiksklks}{\ensuremath{-4.5}\xspace}
\providecommand{\BRphiksklks}{\ensuremath{4.0^{+2.6}_{-2.2}}\xspace}
\providecommand{\signfphiksklks}{\ensuremath{2.2}\xspace}
\providecommand{\fitBiasKsKsKl}{\ensuremath{+0.3}\xspace}
\providecommand{\fitBiasSysKsKsKl}{\ensuremath{0.2}\xspace}
\providecommand{\yieldSystKsKsKl}{\ensuremath{6}\xspace}
\providecommand{\totAddSystKsKsKl}{\ensuremath{5.9}\xspace}
\providecommand{\totMultSystKsKsKl}{\ensuremath{10.6}\xspace}
\providecommand{\totSystKsKsKl}{\ensuremath{0.6}\xspace}
\providecommand{\BRKsKsKl}%
               {\ensuremath{2.4^{+2.7}_{-2.5}\pm\totSystKsKsKl}\xspace}
\providecommand{\signfStatKsKsKl}{\ensuremath{1.0}\xspace}
\providecommand{\signfKsKsKl}{\ensuremath{0.9}\xspace}
\providecommand{\ULStatKsKsKl}{\ensuremath{6.3}\xspace}
\providecommand{\ULKsKsKl}{\ensuremath{7.4}\xspace}
\providecommand{\ULKsKsKlExtreme}{\ensuremath{16}\xspace}

\long\def\inst#1{\par\nobreak\kern 4pt\nobreak
    {\it #1}\par\vskip 10pt plus 3pt minus 3pt}

\graphicspath{%
{./figures/}%
{.}
}

\begin{document}
{

\begin{flushleft}
{\small
\babar-PUB-\BABARPubYear/\BABARPubNumber,  \ SLAC-PUB-\SLACPubNumber \ Rev. Aug. 2006 \\
hep-ex/\LANLNumber  \ \ \ \ \ Phys. Rev. {\bf D74}, 032005 (2006) \\
}
\end{flushleft}

\title{ Search for the decay {\boldmath $B^0 \to \KS \KS \KL$} }

%
\author{B.~Aubert}
\author{R.~Barate}
\author{M.~Bona}
\author{D.~Boutigny}
\author{F.~Couderc}
\author{Y.~Karyotakis}
\author{J.~P.~Lees}
\author{V.~Poireau}
\author{V.~Tisserand}
\author{A.~Zghiche}
\affiliation{Laboratoire de Physique des Particules, F-74941 Annecy-le-Vieux, France }
% ...
\collaboration{The \babar\ Collaboration}
\noaffiliation

\begin{abstract}

\noindent
   We present the first search for the decay $B^0 \to \KS \KS \KL$
   using a data sample of 232 million \BB pairs.
   We find no statistically significant evidence for
   the non-resonant component of this decay.
   Our central value for the branching fraction, assuming
   the true Dalitz distribution is uniform
   and excluding the $\phi$ resonance,
   is ${\cal B}(\Bz \rightarrow \KS \KS \KL) = (\BRKsKsKl) \times 10^{-6}$
   where the errors are statistical and systematic, respectively.
   We set a single-sided Bayesian upper limit of
   ${\cal B}(\Bz \rightarrow \KS \KS \KL) < \ULKsKsKl\times 10^{-6}$
   at 90\% confidence level using a uniform prior probability for
   physical values.
   Assuming the worst-case true Dalitz distribution, where the signal is entirely
   in the region of lowest efficiency,
   the 90\% confidence level upper limit is
   ${\cal B}(\Bz \rightarrow \KS \KS \KL) < \ULKsKsKlExtreme \times 10^{-6}$.
\end{abstract}

\maketitle

\section{INTRODUCTION}
\label{sec:Introduction}

    Measurements from the \babar \ and Belle experiments have
  confirmed the
  Cabibbo-Kobayashi-Maskawa quark mixing matrix (CKM)
  mechanism~\cite{ckm}
  as the dominant source of $CP$
  violation in flavor-changing weak interactions~\cite{kirkbyAndNir}.
     In the Standard Model,
  the mixing-induced, time-dependent $CP$ asymmetry in \Bz
  decays from $b \to s \bar q q$ penguin transitions
  should be the same as the precisely measured $CP$ asymmetry
  in charmonium-\Kz decays, namely $\sin 2\beta$,
  to within a few percent.
  New physics from higher mass scales may contribute to the
  loop in the penguin diagram, which could significantly
  alter the $CP$ asymmetry in penguin-dominated $B$ decays~\cite{grossmanNP}.
  Initial $CP$ asymmetry measurements in $b\to s \bar q q$
  penguin $B$ decays have suggested a possible
  violation of this test of the Standard Model~\cite{btospengpapers},\cite{3ks},\cite{hfag}.

     The $b\to s \bar q q$ penguin decays fall into two categories.
  If the $\bar q q$ can be $\bar u u$, a CKM-suppressed
  tree-level $b\to u$ transition can contribute to the decay in addition
  to the dominant $b\to s \bar q q$ penguin.
  This introduces some uncertainty in the Standard Model prediction
  of the $CP$ asymmetry, since the $b \to u$ and the penguin
  amplitudes have different weak phases.
  On the other hand, decays such as 
  $\Bz \to \KS \KS \KS$~\cite{3ks} and
  $\Bz \to \KS \KS \KL$
  are purely $b\to s \bar s s$ penguin transitions and they can only
  include $b\to u$ decay amplitudes through rescattering,
  thus the Standard Model uncertainty on the predicted $CP$ asymmetry
  for these decays is generally smaller~\cite{soni}.

  It has been noted that three-body \Bz decays of the form $\Bz \to PPP'$,
  where $P$ and $P'$ are spin-0 $CP$ eigenstate neutral particles, are $CP$
  eigenstates\cite{gershon3body},
  thus the $CP$ asymmetry in $\Bz \to \KS \KS \KL$ is not
  diluted, as is generally the case in three-body $\Bz$ decays.
  The $\Bz \to \KS \KS \KL$ would be a valuable addition to
  understanding the $b\to s \bar q q$ penguin $CP$ asymmetry anomaly, if the
  branching fraction is large enough.
  The resonant $\phi \KS$ contribution to this final state,
  neglecting interference effects, would not yield a sample of
  signal events large enough to make an interesting  $CP$ asymmetry
  measurement.
        In the analysis described below, the expected and
        observed number of $\KS \KS \KL$ events
        with a $\KS \KL$ invariant mass in the region
        of the $\phi$ resonance is about eight events.
    Prior to our search, the non-resonant component of the $\KS \KS \KL$
    final state had not been experimentally investigated.  This
    search was motivated by the possibility of discovering a large
    non-resonant contribution to the $\KS \KS \KL$ final state.
    Large non-resonant S-wave amplitudes have been found in
    the Dalitz amplitude analysis of
    $B^+\rightarrow K^+ K^- K^+$~\cite{kpkmkp}
    and $B^0\rightarrow K^+ K^- \kshort$~\cite{kpkmk0}.
  The branching fraction was recently estimated to be~\cite{soni}
  ${\cal B}(\Bz \rightarrow \KS \KS \KL) = \left(5.23^{+2.52 +6.86 +0.05}_{-1.96 -2.53 -0.06}\right) \times 10^{-6}$
  including the $\phi$ resonance.
  The difference between the upper and lower limits of this estimate
  is substantial.
  Another prediction, based on isospin and Bose symmetry~\cite{GandR},
  gives a somewhat small branching fraction of
  ${\cal B}(\Bz \rightarrow \KS \KS \KL) = (1/3) \
   {\cal B}(\Bz \rightarrow \KS \KS \KS) \approx 2 \times 10^{-6}$.


\section{THE \babar\ DETECTOR AND DATASET}
\label{sec:babar}

    The results presented here are based on data collected with the \babar\
    detector~\cite{BABARNIM} at the PEP-II asymmetric $e^+e^-$
    collider~\cite{pep} located at the Stanford Linear Accelerator Center.
    An integrated luminosity of 211 \invfb, corresponding to
    $232\times 10^6$ \BB\ pairs, was recorded at the
    $\Upsilon (4S)$ resonance (center-of-mass energy $\sqrt{s}=10.58\gev$).

    Charged particles from the \epem\ interactions are detected and their
    momenta are measured using five layers of double-sided
    silicon microstrip detectors and a 40-layer drift chamber,
    both operating in the 1.5-T magnetic field of a superconducting
    solenoid. Photons, electrons, and hadronic showers from \KL interactions
    are identified with a CsI(Tl) electromagnetic calorimeter (EMC).
    Further charged particle identification is provided by the
    specific ionization ($dE/dx$) in the tracking devices and by an
    internally reflecting ring imaging Cherenkov detector covering
    the central region.
    The steel of the magnetic-flux return for the
    superconducting solenoid is instrumented
    with resistive plate chambers
    (instrumented flux return, or IFR), which are used to identify muons
    and hadronic showers from \KL interactions.


\section{ANALYSIS METHOD}
\label{sec:Analysis}

  \subsection{Daughter candidate selection}


   We reconstruct \KS candidates through the decay $\KS \to \pi^+ \pi^-$
   only.
   We begin by forming all oppositely-charged combinations of
   reconstructed tracks in the event.
   The invariant mass is required to be in the range of 490 to 506 \mevcc,
   assuming the tracks are pions.
   In addition, the $\chi^2$ probability of the \KS vertex fit
   must be greater than 1\% and the transverse component of the decay
   length is required to be greater than 8 mm.
   Finally, the angle between the \KS flight direction and
   the \KS momentum vector must be less than 0.2 radians.


   We reconstruct \KL candidates by identifying clusters of energy
   in the EMC and hits in the IFR that are isolated from all charged
   tracks in the event.
      The \KL candidates based on IFR clusters must
   have hits in at least two of the 16 to 19 layers in the detector.
   The \KL candidates detected by a hadronic shower in the EMC
   must have a calorimeter energy of at least 200\mev, where
   this energy is from interpreting the calorimeter signal as
   an electromagnetic shower.
   Clusters in the EMC that are consistent with photons from \piz
   decay are vetoed by two methods.
   If the \KL candidate EMC cluster forms an invariant mass
   in the range  of 100 to 150 \mevcc with another EMC cluster with
   a calorimeter energy of at least 100\mev, it is rejected.
      If the \KL candidate cluster has a calorimeter energy greater
      than 1.0\gev and
      two local maxima (or two ``bumps'') in the spatial distribution
      of the energy within the cluster, it may be from a high-energy $\pi^0$, where
      the electromagnetic showers from the two photons are merged into
      one cluster.
      If the two bumps within the cluster form an invariant mass
      greater than 110\mevcc, the candidate is rejected.

      Further background rejection for EMC \KL candidates
   is achieved by using a neural network trained on
   signal, $q\bar q$ continuum
   (where $q=u,d,s,c$),
   and $B\overline{B}$
   Monte Carlo samples to distinguish \KL clusters from fake clusters.
   The neural network inputs are the EMC cluster energy and
   the following six shower-shape variables:
   \begin{itemize}
     \item The lateral moment LAT of the shower energy deposition~\cite{latMoment}
           defined as
           LAT~$=\sum_{i=3}^n E_i r_i^2 /
           (E_1 r_0^2 + E_2 r_0^2 + \sum_{i=3}^n E_i r_i^2)$
           where the $n$ crystals in the EMC cluster are ranked in order of
           deposited energy ($E_i$) , $r_0 = 5$~cm is the average distance
           between crystal centers, and $r_i$ is the radial distance of crystal
           $i$ from the cluster center.
     \item The second radial moment of the shower energy deposition,
          defined as $\sum_i E_i r^2_i$ where
          $r_i$ is the radial distance of crystal $i$ from the cluster center.
      \item The energy sum
           of a $3\times 3$ block of crystals, centered on the single crystal with the
           most energy, divided by the larger 
           $5\times 5$ block, also centered in the
           same way.
      \item The energy of the
           most energetic crystal in the cluster divided by the energy sum of
           the $3\times 3$
           crystal block with the most energetic crystal in the center.
      \item The Zernike moments~\cite{zern} $A_{2,0}$ and $A_{4,2}$ defined
           below.
   \end{itemize}
   The Zernike moment $A_{n,m}$ is defined as
   \[
       A_{n,m} = \left| \sum_i \, \frac{E_i}{E_{\rm tot}} \,
             f_{n,m}(r_i/R_0) \, e^{i m \phi_i} \right|
   \]
    with
    \[
         f_{2,0}(x) = 2 x^2 - 1 \ \ \ \ \ \ {\rm and} \ \ \ \ \ \ \ f_{4,2}(x) = 4 x^4 - 3 x^2
    \]
    where $r_i$ and $\phi_i$ are the radial and angular
    separation of crystal $i$
    with respect to the cluster center,
    $E_{\rm tot}$ is the total cluster energy,
    and $R_0$ is a cutoff radius of 15~cm.
   Only \KL candidates that pass an optimized cut on the neural
   network output are retained.
   This cut has a signal efficiency of 85\% and rejects
   70\% of the EMC fake \KL background.

  \subsection{{\boldmath \Bz} candidate selection}


   We reconstruct \Bz candidates from selected \KL clusters
   and pairs of \KS candidates that do not share any tracks.
   We require the sum of the \KS momentum magnitudes in
   the center-of-mass frame
   to be at least 2.1\gevc, which ensures consistency with
   the kinematics of a $\Bz \to KKK$ decay.
   Only the direction of the \KL is reconstructed, from the vector
   defined by the primary vertex and the center of the neutral
   cluster.
   We compute the \KL momentum by constraining the
   $\KS \KS \KL$ invariant mass to be the known mass
   of the \Bz\cite{PDG2004}.
       We reject fake \KL candidates by using the difference between
       the  \KL transverse momentum, computed from the $\Bz$ mass
       constraint, and the transverse momentum along the \KL direction
       that is missing from the event.
       The reconstructed event missing transverse momentum is
       calculated without using the \KL cluster in the momentum sum
       and projected along the computed direction of the \KL.
       This missing transverse momentum difference (reconstructed
       minus calculated) is required to be greater than $-0.5\gevc$.
        This requirement and the previously mentioned requirement
        on the EMC \klong neural network output were simultaneously
        optimized to give the greatest signal significance
        $S/\sqrt{S+B}$, where $S$ and $B$ are the expected number of
        signal and background events, assuming a signal branching
        fraction of $5\times 10^{-6}$~\cite{soni}.

   The difference in energy $\Delta E$ between the reconstructed
   \Bz candidate and the beam energy in the center-of-mass frame
   is the main variable used to distinguish properly reconstructed
   signal events from combinatoric background.
      The missing transverse momentum difference distributions for
      the signal Monte Carlo sample and the background-dominated $\Delta E$
      sideband ($\Delta E > 0.010$~GeV in data)
      are shown in Figure~\ref{fig:ptmiss}.

   \begin{figure}
      \begin{center}
      \epsfig{file=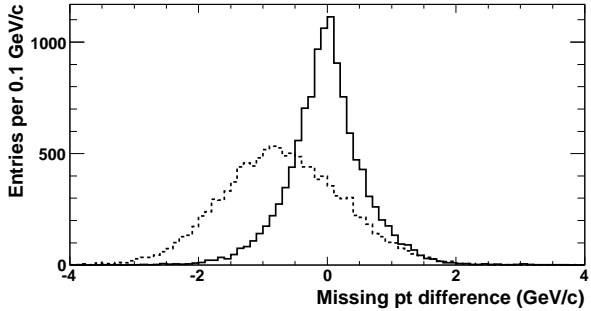,width=0.99\linewidth}
      \caption{
         Missing transverse momentum difference distributions for
         the signal Monte Carlo sample (solid) and the background-dominated
         $\Delta E$ sideband ($\Delta E > 0.010$~GeV)
         after all other selection criteria (see text)
         have been applied.
         The signal Monte Carlo sample distribution
         has been arbitrarily normalized
         to the number of entries in the sideband distribution.
      }
      \label{fig:ptmiss}
      \end{center}
   \end{figure}


   We
   distinguish the non-resonant three-body \Bz decay
   from two-body \Bz decays to the same final state in this
   analysis.
   There are four two-body \Bz decays through charmonium
   that can go to the final state $\KS \KS \KL$, namely:
   $J/\psi \KS$, $\chi_{c0} \KL$, $\chi_{c2} \KL$, and
   $\psi(2S) \KS$.
   These are particularly unwanted, since they are from
   a color-suppressed tree decay amplitude, not
   the $b\rightarrow s \bar q q$ penguin decay amplitude.
   The $\chi_c$ modes have unknown \Bz branching fractions.
   We veto \Bz candidates consistent with $\Bz \to \chi_{c0} \KL$
   or $\Bz \to \chi_{c2} \KL$, where the $\chi_c$ decays to
   $\KS \KS$, by removing candidates
   with a $\KS \KS$ invariant mass in the range
   3.400 to 3.429 \gevcc or
   3.540 to 3.585 \gevcc respectively.
   The combined \Bz and daughter branching fractions for
   the $J/\psi \KS$ and $\psi(2S) \KS$ modes are 0.062 and
   0.016 times $10^{-6}$ respectively.
   We expect about two events from two-body \Bz decays through charmonium
   in our final sample.

   We also remove \Bz candidates consistent with
   $\Bz \to \phi \KS$ with $\phi \to \KS \KL$ by requiring
   the invariant mass of both $\KS \KL$ combinations to be above 
   1.049\gevcc, though as a cross check, we measure the
   branching fraction in only the $\phi \KS$ region, where
   one $\KS \KL$ combination has an invariant mass less than
   1.049\gevcc.
   These two-body \Bz decay vetos are 96.6\% efficient for
   the $\Bz \to \KS \KS \KL$ signal Monte Carlo sample
   generated with
   a uniform true Dalitz distribution.

  \subsection{ Event selection }


   The main source of background is from random $\KS \KS \KL$
   combinations from continuum $q \bar q$
   events.
   We combine 3 variables in a neural network that is trained
   using Monte Carlo samples to distinguish signal from continuum events.
   The first input is the cosine of the angle of the
   $\Bz$ momentum with respect to the beam axis
   in the center-of-mass frame $\theta^*_B$,
   which is flat
   for continuum background whereas the
   signal probability is proportional to $\sin^2 \theta^*_B$.
   The other two inputs are topological variables that are
   commonly used to distinguish jet-like continuum events
   from the more isotropic particle distributions in \BB
   events.
   The first is the cosine of the angle $\theta_T$ between the
   thrust axis of the \Bz candidate in the center-of-mass
   frame and that of the rest of the charged tracks and neutral calorimeter
   clusters in the event.  The distribution of $|\cos \theta_T|$ is sharply
   peaked near $1.0$ for combinations drawn from jet-like $q\bar q$
   pairs, and nearly uniform for
   \Bz meson
   decays.
   The second is a linear combination of the zeroth and second
   angular moments $L_{0,2}$ of the energy flow about the \Bz
   thrust axis.
   The moments are defined by $ L_j = \sum_i
   p_i\times\left|\cos\theta_i\right|^j,$ where $\theta_i$ is the angle
   with respect to the $\Bz$ thrust axis of track or neutral cluster $i$,
   $p_i$ is its momentum, and the sum excludes the $\Bz$ candidate daughters.
      Distributions of the neural network output $NN$ for
      the signal Monte Carlo sample and the background-dominated
      $\Delta E$ sideband
      are shown
      in Figure~\ref{fig:esnn}.
   We require $NN>0.5$ to remove events that have little probability
   of being signal.

   \begin{figure}
      \begin{center}
      \epsfig{file=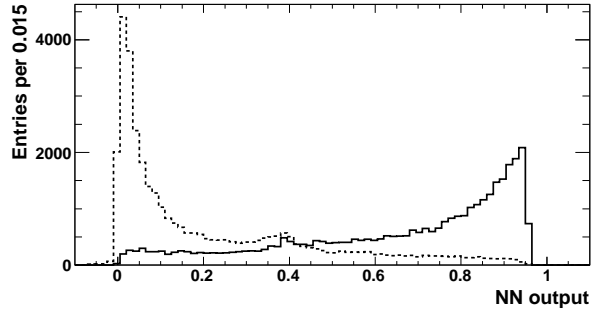,width=0.99\linewidth}
      \caption{
         Neural network output ($NN$)
         distributions for the signal Monte Carlo sample (solid)
         and the background-dominated
         $\Delta E$ sideband ($\Delta E > 0.010$~GeV)
         after all other selection criteria (see text)
         have been applied.
         The signal Monte Carlo sample distribution
         has been arbitrarily normalized
         to the number of entries in the sideband distribution.
      }
      \label{fig:esnn}
      \end{center}
   \end{figure}


   After all of the requirements stated thus far, if an event
   has more than one \Bz candidate, we choose the best one
   by selecting on the quality of the \KL cluster.
   If there are one or more EMC \KL candidates, the best \KL candidate is
   the one with the highest cluster calorimeter energy.
   If there are no EMC \KL candidates, the best \KL candidate
   is the one with the highest number of IFR layers with hits in the
   \KL cluster.
   If there is more than one \KS pair that uses the same
   (best) \KL cluster, the best \Bz candidate is the one with
   the lowest \KS mass $\chi^2$, defined
   below
   \[ \chi^2 = \left(\frac{\delta m_1}{\sigma}\right)^2
             + \left(\frac{\delta m_2}{\sigma}\right)^2
             \]
   where $\delta m_i$ is the difference between the reconstructed
   invariant mass of \KS candidate $i$ and the known $\KS$ mass~\cite{PDG2004}
   and $\sigma$ is the invariant mass resolution (2.9~MeV/c$^2$).

   Our final analysis sample contains 5892 events with $\Delta E$
   in the range of $-0.010\gev$ to $0.080\gev$.
   Signal events appear mostly in the range of $-0.010\gev$ to
   $0.010\gev$.
   The $\Delta E>0.010$\gev region is dominated by combinatoric
   background.
   The signal efficiency varies from 4\% to 14\% depending on
   the position in the Dalitz plot.
   The signal efficiency as a function of Dalitz plot position
   and a histogram of the reconstruction efficiency for 322
   uniformly distributed points in the Dalitz plot are shown
   in Figures~\ref{fig:eff2D} and~\ref{fig:eff} respectively.
   For a uniform true Dalitz distribution,
   the average signal efficiency
   is $\kskskslCorrEff$\%. 

      \begin{figure}[h]
      \begin{center}
      \epsfig{file=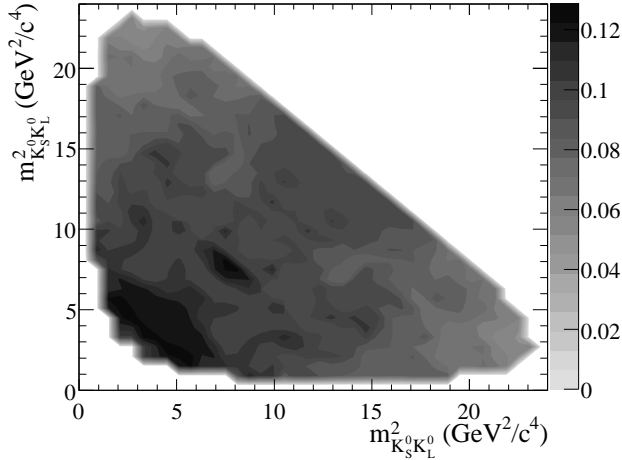,width=0.95\linewidth}
      \caption{
         The reconstruction efficiency for signal
         as a function of
         position in the symmetrized Dalitz plot.
      }
      \label{fig:eff2D}
      \end{center}
      \end{figure}
      \begin{figure}[h]
      \begin{center}
      \epsfig{file=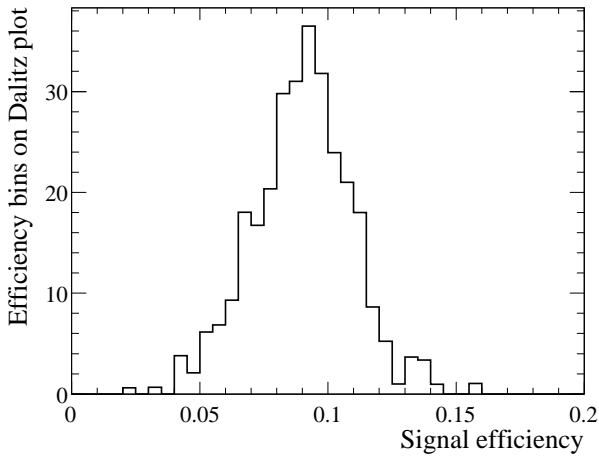,width=0.95\linewidth}
      \caption{
         A histogram of the reconstruction efficiency for signal
         for 322 uniformly distributed points in the Dalitz
         plot.
      }
      \label{fig:eff}
      \end{center}
      \end{figure}


   \subsection{ {\boldmath $J/\psi \KL$} control sample }

   We use a sample of $\Bz \to J/\psi \KL$ decays from the data
   to calibrate the reconstruction and selection efficiency and
   the $\Delta E$ resolution.
   The $J/\psi$ is reconstructed in the $e^+e^-$ and $\mu^+\mu^-$
   channels.
   We apply the same \KL
   selection,
   projected missing transverse momentum difference,
   and $NN$
   criteria as for our $\KS \KS \KL$ selection.
        The ratio of the $\KS \KS \KL$ and $J/\psi \KL$ Monte Carlo
        efficiencies for these criteria is
        $\epsilon_{\KS \KS \KL}/\epsilon_{J/\psi \KL} = 0.96$.
   We compare the number of fitted $J/\psi \KL$ events to the
   predicted number of events, based on the known branching
   fractions and the Monte Carlo efficiency.
        The $J/\psi \KL$ reconstruction efficiency for all selection
        criteria is 12.5\%.
   We find $N_{\rm obs} = 1420 \pm 56$ $J/\psi \KL$ events,
   consistent with the predicted yield
   of $N_0 = 1426 \pm 86$. We correct the $\KS \KS \KL$ efficiency
   by multiplying the MC efficiency by a
   correction factor of $0.96 \pm 0.08$,
   which is defined as $(N_{\rm obs}/N_0)\times(\epsilon_{\KS \KS \KL}/
   \epsilon_{J/\psi \KL})$.
   The uncertainty on the correction factor
   includes the uncertainties on the
   relevant branching fractions for $J/\psi \KL$, the
   statistical error from the $J/\psi \KL$ fit, and our
   estimated uncertainty on relating the
   $J/\psi \KL$ and $\KS \KS \KL$ selection and
   reconstruction efficiencies.

   The $\Delta E$ resolution is better for EMC candidates because
   the position of the hadronic shower is more precisely measured.
   We also use the $J/\psi \KL$ to measure the signal $\Delta E$
   resolution separately for \KL candidates reconstructed in the
   EMC and IFR.
   Figure~\ref{fig:jpsikl_deProj} shows the fitted $\Delta E$
   distributions for EMC and IFR \KL candidates.

      \begin{figure}[h]
      \begin{center}
      \epsfig{file=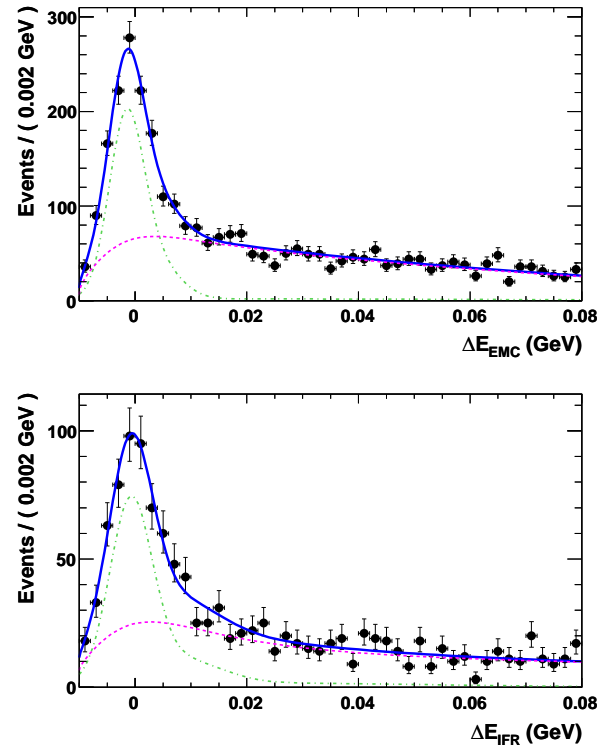,width=0.95\linewidth}
      \caption{
         Fitted $\Delta E$ distributions
         of the $J/\psi \KL$ control sample
         for EMC (top) and IFR (bottom) \KL candidates.
         The points with error bars are histograms of the data sample.
         The solid 
         curve is the total PDF.
         The dot-dashed 
         curve is the signal PDF and the
         dashed 
         curve is the background PDF.
      }
      \label{fig:jpsikl_deProj}
      \end{center}
      \end{figure}
%


   \subsection{ Signal yield determination }

   We use an extended unbinned maximum likelihood fit to determine the
   number of signal events in our final
   sample.
   The likelihood for an event is the product of probability
   density functions (PDFs) for the two main discriminating variables
   in the analysis: $\Delta E$ and $NN$.
   Separate PDFs are used for EMC and IFR candidates due to the
   difference in $\Delta E$ resolution and the fact that some
   background channels mostly produce fake signal for EMC
   \KL candidates only.
       The EMC and IFR $\KL$ samples are fitted simultaneously
       and the relative fraction of EMC signal events is
       taken from the $J/\psi \KL$ control sample.
   Separate PDFs are also used for signal, combinatoric background,
   and three classes of background from $B$ decays similar to
   our signal (``peaking backgrounds'' described below
   in Section~\ref{sec:peakingbg}).
     The combinatoric background
     includes random $\kshort \kshort \klong$ combinations from
     \BBbar events  as well as from continuum events.

         Figure~\ref{fig:PDF_de_esNN} shows the $\Delta E$ and
         $NN$ PDFs for all five fit components.
   The numbers of signal and combinatoric background events
   are determined from the fit.
   The peaking background yields are fixed to values based
   on known and estimated branching fractions and then
   varied in the evaluation of systematic uncertainties.

   The functional form of the signal PDF for $\Delta E$ is
   a triple Gaussian distribution.
   The mean and width of the core Gaussian  distribution are determined
   separately for EMC and IFR \KL candidates from the
   $J/\psi \KL$ control sample and held fixed in
   the $\KS \KS \KL$ fit.
   The remaining Gaussian parameters for the signal
   $\Delta E$ PDF are determined
   from the signal Monte Carlo sample and held fixed.
      The signal PDF for $NN$ is a 4th-order polynomial.
      The polynomial coefficients are determined from the signal
      Monte Carlo sample and held fixed in the fit.
      The combinatoric background $\Delta E$ PDF is an
      ARGUS function~\cite{argus-function}.
      The $NN$ PDF for combinatoric background is the sum of
      a 1st-order polynomial and an ARGUS function.
   The PDF shape parameters for the combinatoric background
   component are free in the fit.

  \subsection{Backgrounds from other {\boldmath $B$} decays}
  \label{sec:peakingbg}


   Backgrounds from $B$ decays to final states similar to
   $\KS \KS \KL$ can look similar to our signal in our discriminating
   variables $\Delta E$ and $NN$.
   We call events from these decays ``peaking backgrounds.''
   The largest single source of peaking background is the
   decay $\Bz \to \KS \KS \KS$.  One of the \KS can decay
   to $\pi^0\pi^0$, where one or more of the photons from
   either $\pi^0$ fakes the \KL cluster in the EMC.
   The world average $\KS \KS \KS$ branching fraction
   from the Heavy Flavor Averaging Group (HFAG)\cite{hfag},\cite{3ks}
   is ${\cal B}(\Bz \rightarrow \KS \KS \KS) = (6.2 \pm 0.9) \times 10^{-6}$.
   We expect 20 $\KS \KS \KS$ events in our fit sample and
   vary this number by $\pm 5$ events in the evaluation
   of the systematic errors.

   Decays of the type $KKK^*$ can produce a $\KS \KS \KL \pi$
   final state, which can look similar to our signal if the
   momentum of the additional $\pi$ is low.
   The branching fractions for the relevant $KKK^*$ have not
   been measured.
       We assume they are each of the same order as our expected
       signal, based on comparing the relative branching ratios
       of $B\rightarrow K \pi$ decays with $B\rightarrow K^*\pi$
       and $B\rightarrow K \rho$ decays.
   We estimate a total branching fraction of $30\times 10^{-6}$
   for this inclusive final state and vary this assumption by
   $\pm 15 \times 10^{-6}$ in the evaluation of systematic errors.
   This gives us an expected $70 \pm 35$ events from $KKK^*$ in the
   fit sample.
       In the evaluation of the 90\% C. L. upper limit on the
       branching fraction, we set the $KKK^*$ event yield to
       zero.  This is the most conservative assumption, since
       the signal and $KKK^*$ event yields are anti-correlated
       in the fit.

   Finally, the decay $\Bz \to \KS \KS \pi^0$ can also mimic
   our signal if the $\pi^0$ is misidentified as a \KL in
   the EMC.
   The branching fraction for this mode has also not been measured.
        We estimate the branching fraction, relative to our signal, by
        assuming that the tree-to-penguin ratio for the three-body decays
        is the same as for $B^0\rightarrow \pi^0\pi^0$ vs $B^0\rightarrow K^0\pi^0$
        and that our signal branching fraction is about $5 \times 10^{-6}$~\cite{soni}.
   Our estimate for the branching fraction is
   ${\cal B}(\Bz \rightarrow \KS \KS \pi^0) \approx 1.6 \times 10^{-6}$
   and we vary this assumed branching fraction in the range
   $[0.1,5.0]\times 10^{-6}$ in the evaluation of systematic errors.
   This gives us an expected
   $2 ^{+4}_{-2}$ events
   from
   $\KS \KS \pi^0$ in the fit sample.

   The $\Delta E$ and $NN$ PDFs for the three peaking background
   components above are determined from Monte Carlo samples.
   In general, the $\Delta E$ shape peaks near $\Delta E=0$,
   though there is a substantial tail that extends to high
   $\Delta E$ values.
      We use the sum of an ARGUS function and a Landau
      function for the peaking background $\Delta E$ PDF.
   The $NN$ shape is quite similar to the signal shape for
   all three peaking backgrounds.
      The form of the $NN$ PDF is a polynomial.
   The $\Delta E$ and $NN$ PDFs for the three peaking background
   components are shown in Figure~\ref{fig:PDF_de_esNN}:
   c, d, and e.

      \begin{figure}[h]
      \begin{center}
      \includegraphics[%
      width=\linewidth%
      ]{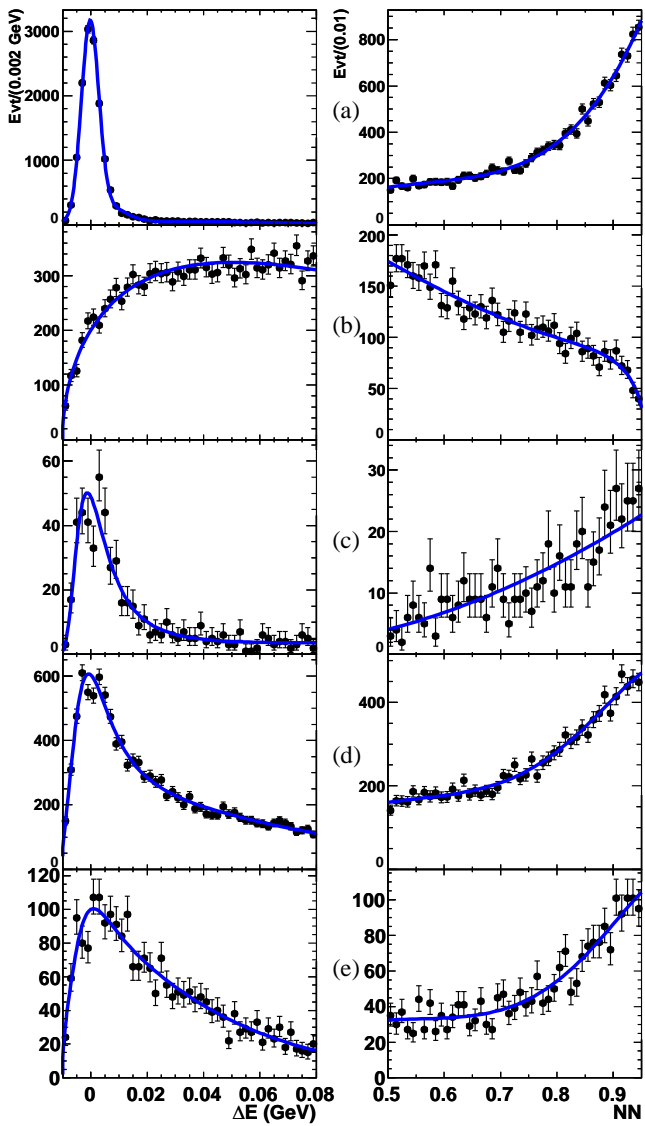}%
      \caption{
         Distributions of $\Delta E$ (left) and $NN$ (right)
         for the five components of the fit: (a) \kskskl signal,
         (b) combinatoric background, (c) $\KS \KS \pi^0$,
         (d) $\KS \KS \KS$, and (e) $KKK^*$.
         The combinatoric $NN$ distribution is from the
         data sideband ($\Delta E>0.010$\gev).
         The rest of the distributions are from Monte Carlo
         samples.
      }
      \label{fig:PDF_de_esNN}
      \end{center}
      \end{figure}


   \section{RESULTS}

   Table~\ref{tab:yResults} lists the results of the maximum likelihood
   fit.
   We find $\YKsKsKl^{\PErrKsKsKl}_{\MErrKsKsKl} \pm \yieldSystKsKsKl$ events
   for the $\KS \KS \KL$ signal yield, where the first errors are statistical
   and the second error is systematic.
   %
   %
   The maximum likelihood fit bias of +0.3 signal events
   was evaluated from an ensemble of data sets composed of
   fully simulated signal and $B$ background Monte Carlo events.
   The combinatoric background for these datasets was generated
   from the PDF parameters of the fit to the data.
   This average bias of 0.3 events and the expected 2.1 events
   from \Bz decays through charmonium are
   subtracted from the fitted signal yield.

   The systematic errors on the fitted signal
   yield and the signal branching fraction
   are listed in
   Table~\ref{tab:ySyst}.
   %
   %
   The additive contributions to the systematic error
   come from the uncertainty on our estimation of
   the $KKK^*$ component normalization (5.2 events),
   varying the yields of the other fixed peaking background
   components and the fixed PDF parameters (2.6 events),
   our uncertainty on the fit bias correction (0.2 events),
   and the uncertainty on the charmonium background subtraction (1.1 events).
   The multiplicative contributions are from
   the uncertainties on the \kshort and \klong reconstruction
   efficiencies (6.8 and 8.0\%), the number of \BB events in the
   sample (1.1\%), and the $\kshort \to \pi^+ \pi^-$ branching fraction.

   Figure~\ref{fig:sigProj} shows the fitted distributions of $\Delta E$
   and $NN$, where a cut has been made on the fit variable not shown
   (e.g. there is a $NN$ cut applied for the $\Delta E$ plot)
   to enhance the signal.
   The signal efficiency of this cut is 57\% for the $\Delta E$ distribution
   and 71\% for the $NN$ distribution.

      \begin{figure}[h]
      \begin{center}
      \epsfig{file=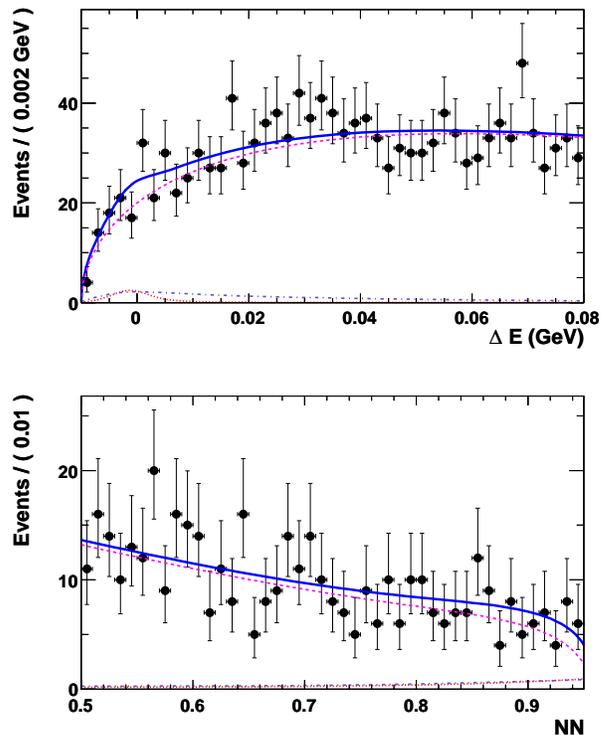,width=0.95\linewidth}
      \caption{
         Distributions of $\Delta E$ (top) and $NN$ (bottom)
         for the \kskskl fit.
         The plot of $\Delta E$ ($NN$) is for events
         passing a cut on $NN$ ($\Delta E$) which enhances the
         signal.
   The signal efficiency of this cut is 57\% for the $\Delta E$ distribution
   and 71\% for the $NN$ distribution.
         The points with error bars are histograms of data samples.
         The solid 
          curves are total PDFs.
         The 
          dashed curves are combinatoric backgrounds.
         The dot-dashed curves are peaking \B backgrounds.
         The 
         dotted curves are signal PDFs.
      }
      \label{fig:sigProj}
      \end{center}
      \end{figure}

   The true distribution of $\KS \KS \KL$ events in the Dalitz plot
   is unknown.
   We use the average signal efficiency
   assuming a uniform true Dalitz distribution
   (\kskskslCorrEff \%)
   to compute
   the $\KS \KS \KL$ branching fraction.
      We found no significant dependence of the signal $\Delta E$
      resolution or the signal $NN$ shape on the Dalitz plot variables.
   With these assumptions, we find a \kskskl branching fraction,
   excluding the $\phi$ resonance, of
   ${\cal B}(\Bz \rightarrow \kskskl) = (\BRKsKsKl) \times 10^{-6}$, where the first
   error is statistical and the second error is systematic.
   The dominant systematic error is from
   the uncertainties of peaking \B backgrounds (23\% relative).

   Figure~\ref{fig:NLL} shows a scan of the negative log likelihood
   as a function of the \kskskl branching fraction, where the minimum
   negative log likelihood has been subtracted.
           In order to remove any dependence on our estimate of
           the total $KKK^*$ branching fraction, we conservatively
           fix the $KKK^*$ yield to zero in the scan of the log
           likelihood.
   We compute a one-sided Bayesian 90\% confidence-level upper limit on
   the branching fraction assuming a uniform prior probability for
   positive (physical) branching fraction values.
   Systematic errors are included by convolving the fit likelihood
   with a Gaussian distribution with a width corresponding to the
   total systematic error, excluding the uncertainty on the
   $KKK^*$ yield, since it is fixed to zero in the likelihood scan.
   The result for the non-resonant three-body branching fraction is
   ${\cal B}(\Bz \rightarrow \kskskl) < \ULKsKsKl \times 10^{-6}$ at 90\% C.L.
   Assuming the worst-case true Dalitz distribution, where the signal is entirely
   in the region of lowest efficiency (4\%),
   the 90\% C. L. upper limit on
   the branching fraction is $\ULKsKsKlExtreme\times 10^{-6}$.

      \begin{figure}[h]
      \begin{center}
      \epsfig{file=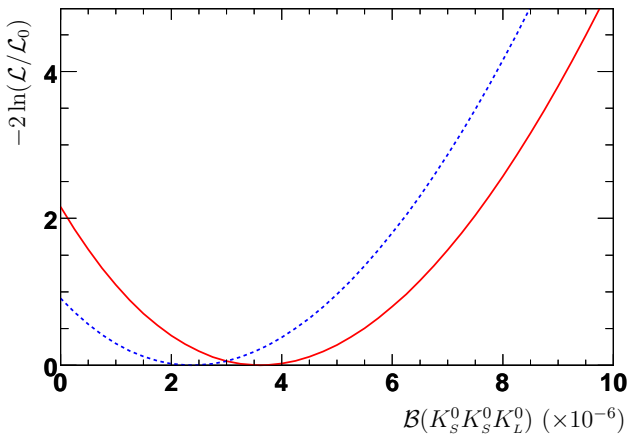,width=0.95\linewidth}
      \caption{
         Scan of the fit negative log likelihood
         as a function of the \kskskl branching fraction, where the minimum
         negative log likelihood has been subtracted.
         A uniform true Dalitz distribution
         has been assumed for
         the \kskskl signal.
         The 
         solid curve includes the systematic uncertainty
         with yield of $KKK^*$ fixed at 0 to calcuate upper limit.
      }
      \label{fig:NLL}
      \end{center}
      \end{figure}


   As a cross check on the analysis, we have performed the fit
   only in the $\phi \KS$ region of the Dalitz plot where
   the invariant
   mass of one of the $\KS \KL$ pairs is less than 1.049\gevcc.
   The results are given in the second column of Table~\ref{tab:yResults}.
   We find $\Yphiksklks^{\PErrphiksklks}_{\MErrphiksklks}$ signal
   events, which corresponds to a branching fraction of
   ${\cal B}(\Bz \rightarrow \phi \KS) = (4.0^{+2.6}_{-2.2})\times 10^{-6}$, where
   the errors are statistical only.
   This is consistent with the world average value of
   $0.5 \cdot {\cal B}(\Bz \rightarrow \phi K^0) = (4.3^{+0.7}_{-0.6})\times 10^{-6}$\cite{PDG2004}.

           We checked our estimation of the $KKK^*$ peaking
           background yield by allowing it to float in the fit.
           This fit gave a $KKK^*$ yield of $-54 \pm 170$ events,
           which is consistent with our estimation of
           $70 \pm 35$ events.

   \begin{table}[htb!]
   \begin{center}
   \caption[Yield fit results]%
     {
     Results of the fit for yields, branching fraction calculation,
     and branching fraction upper limit (UL) calculation.
     The $\KS \KS \KL$ efficiency below assumes
     a uniform true Dalitz distribution.
     The maximum likelihood fit bias and charmonium
     background are subtracted from the signal yield
     branching fraction calculation.
     }
   \label{tab:yResults}
   \bigTableStretch
   \begin{tabular}{lcc}
     \dbline
     Mode                          & \kskskl           & \phiks  \\
     \hline
     Events to fit                 & \NsimEvtKsKsKl    & $210$ \\
     Signal yield   & $23^{+23}_{-22}\pm6$
                         & $\Yphiksklks^{\PErrphiksklks}_{\MErrphiksklks}$ \\
     Comb. Bkg                     & \YBkgKsKsKl       & \YBkgPhiKs \\
     \KKKst Bkg                    & \NsimKKKst (fixed)        & NA \\
     \ksksks Bkg                   & \NsimKsKsKs (fixed)       & NA \\
     \KKPiZ Bkg                    & \NsimKKPiZ (fixed)         & NA \\
     Estimated fit bias (evt)             & \fitBiasKsKsKl            & $0.0$ \\
     $\ccbar K_{S/L}$              & $2.1$             & NA \\
     ~~MC $\epsilon$ (\%)          & \MCeffKsKsKl      & $6.1$ \\
     ~~\kshort corr. (\%)                      & \multicolumn{2}{c}{$96.2$}  \\
     ~~\klong corr. (\%)                       & \multicolumn{2}{c}{\effKL}  \\
     Corr. $\epsilon$ (\%)               & \kskskslCorrEff  & \phiksklksCorrEff \\
     $\prod{\calB_i}$ (\%)                     & $47.5$  & $16.2$ \\
     Corr. $\epsilon\times\prod{\calB_i}$ (\%) & \ksksklEffProdBRs 
                                               & \phiksklksEffProdBRs \\
     \hline
     \calB $(\times 10^{-6})$              & \BRKsKsKl & \BRphiksklks \\
     Stat. signf. ($\sigma$)             & \signfStatKsKsKl  & \signfphiksklks \\
     Signf. w/ syst. ($\sigma$)                & \signfKsKsKl \\
     \hline
     90\% CL UL \calB $(\times10^{-6})$ (stat.)        & \ULStatKsKsKl & NA \\
     90\% CL UL \calB $(\times10^{-6})$ (incl. syst.)  & \ULKsKsKl & NA \\
     \dbline
   \end{tabular}
   \end{center}
   \end{table}

   \begin{table}[htb!]
   \caption{Estimates of systematic errors. Multiplicative systematic
   errors are in percent while additive systematic errors are in events.
   The fit yield systematic error is due to fixed fitting parameters,
   mainly from \kskskl \DE PDF parameters, and the uncertainties of
   fixed peaking \B background yields.
   The fit bias error is one half of the bias.
   The $\ccbar K_{S/L}$ error is the uncertainty of charmonium background
   subtraction,
   estimated as
   one half of the subtraction.
   }
   \label{tab:ySyst}
   \begin{center}
   \bigTableStretch
   \begin{tabular}{lc}
     \dbline
     Quantity                  & \kskskl \\
     \sgline
   Additive errors (events) \\
   ~~$KKK^*$ normalization     &  5.2  \\
   ~~Fit yield                 &  2.6  \\
   ~~Fit bias                  &  \fitBiasSysKsKsKl  \\
   ~~$\ccbar K_{S/L}$          &  $1.1$  \\
   \sgline
   Total additive (events)     &  \totAddSystKsKsKl  \\
   \sgline
   Multiplicative errors (\%) \\
   ~~\KS\ efficiency           &  $6.8$   \\
   ~~\klong efficiency         &  \effKLSyst  \\
   ~~Number \BB\               &  $1.1$   \\
   ~~${\cal B}(\Bz \rightarrow \kshort \to \pi^+ \pi^-)$       &  $0.4$   \\
   \sgline
   Total multiplicative (\%)   &  \totMultSystKsKsKl  \\
   \sgline
   Total errors \lbrack\bfemsix\rbrack  &  \totSystKsKsKl   \\
   \dbline
   \end{tabular}
   \end{center}
   \end{table}


    \section{SUMMARY}

    We have searched for the decay $\Bz \to \kskskl$ using 232 million
    \BB pairs recorded by the \babar \ experiment.
    We find no significant evidence for this decay.
    The central value for the branching fraction,
    assuming a uniform true Dalitz distribution
    for the signal and excluding the $\phi$ resonance,
    is ${\cal B}(\Bz \rightarrow \kskskl) = (\BRKsKsKl) \times 10^{-6}$, where the first
    error is statistical and the second error is systematic.
    This corresponds to a Bayesian 90\% C.L. upper limit of
    ${\cal B}(\Bz \rightarrow \kskskl) < \ULKsKsKl \times 10^{-6}$.
   Assuming the worst-case true Dalitz distribution, where the signal is entirely
   in the region of lowest efficiency,
    the upper limit on the branching fraction is
    $\ULKsKsKlExtreme\times 10^{-6}$.
       Our results show that the \kskskl channel will be of limited
       use in understanding the $b \rightarrow s \bar q q$ penguin
       $CP$ anomaly, due to the low efficiency times branching fraction,
       which limits the yield of signal events.

    \section{ACKNOWLEDGMENTS}
    \label{sec:Acknowledgments}
 %

    We are grateful for the 
extraordinary contributions of our \pep2\ colleagues in
achieving the excellent luminosity and machine conditions
that have made this work possible.
The success of this project also relies critically on the 
expertise and dedication of the computing organizations that 
support \babar.
The collaborating institutions wish to thank 
SLAC for its support and the kind hospitality extended to them. 
This work is supported by the
US Department of Energy
and National Science Foundation, the
Natural Sciences and Engineering Research Council (Canada),
Institute of High Energy Physics (China), the
Commissariat \`a l'Energie Atomique and
Institut National de Physique Nucl\'eaire et de Physique des Particules
(France), the
Bundesministerium f\"ur Bildung und Forschung and
Deutsche Forschungsgemeinschaft
(Germany), the
Istituto Nazionale di Fisica Nucleare (Italy),
the Foundation for Fundamental Research on Matter (The Netherlands),
the Research Council of Norway, the
Ministry of Science and Technology of the Russian Federation, and the
Particle Physics and Astronomy Research Council (United Kingdom). 
Individuals have received support from 
CONACyT (Mexico), the Marie-Curie Intra European Fellowship program (European Union),
the A. P. Sloan Foundation, 
the Research Corporation,
and the Alexander von Humboldt Foundation.

\end{document}